\documentclass{PoS}

\usepackage{amsmath}

\newcommand{\be}{\begin{equation}}
\newcommand{\ee}{\end{equation}}
\newcommand{\bea}{\begin{eqnarray}}
\newcommand{\eea}{\end{eqnarray}}
\newcommand{\bt}{\begin{tabbing}}
\newcommand{\et}{\end{tabbing}}
\newcommand{\bi}{\begin{itemize}}
\newcommand{\ei}{\end{itemize}}
\newcommand{\ben}{\begin{enumerate}}
\newcommand{\een}{\end{enumerate}}

\newcommand{\calO}{{\mathcal O}}

\newcommand{\bfx}{{\bf x}}
\newcommand{\bfr}{{\bf r}}

\title{
   \begin{picture}(0,0)(0,0)%
   \put(355,75){\makebox(0,0)[l]{\textnormal{\normalsize KEK-CP-226}}}%
   \end{picture}%
   Flavor-singlet mesons in $N_f=2+1$ QCD with dynamical overlap quarks
}

\ShortTitle{Flavor-singlet mesons in $N_f\!=\!2\!+\!1$ QCD 
            with dynamical overlap quarks}

\author{
   JLQCD and TWQCD collaborations: 
   \speaker{T.~Kaneko}$^{a,b}$\thanks{E-mail: takashi.kaneko@kek.jp}, 
   S.~Aoki$^{c,d}$, 
   T.~W.~Chiu$^{e}$, 
   H.~Fukaya$^f$, 
   S.~Hashimoto$^{a,b}$, 
   T.~H.~Hsieh$^g$,
   J.~Noaki$^{a}$, 
   E.~Shintani$^{h}$
   and 
   N.~Yamada$^{a,b}$
   \\
   \\
   \\
   \llap{$^a$}
   KEK Theory Center, 
   High Energy Accelerator Research Organization (KEK),
   Ibaraki 305-0801, Japan 
   \\
   \llap{$^b$}
   School of High Energy Accelerator Science,
   The Graduate University for Advanced Studies (Sokendai),
   Ibaraki 305-0801, Japan
   \\
   \llap{$^c$}
   Graduate School of Pure and Applied Sciences, 
   University of Tsukuba, Ibaraki 305-8571, Japan
   \\
   \llap{$^d$}
   Riken BNL Research Center, 
   Brookhaven National Laboratory, Upton, New York 11973, USA
   \\
   \llap{$^e$}
   Physics Department, Center for Theoretical Sciences,
   and Center for Quantum Science and Engineering,
   National Taiwan University, Taipei, 10617, Taiwan
   \\
   \llap{$^f$}
   Department of Physics, Nagoya University, 
   Nagoya 464-8602, Japan
   \\
   \llap{$^g$}
   Research Center for Applied Sciences,
   Academia Sinica, Taipei 115, Taiwan
   \\ 
   \llap{$^h$}  
   Department of Physics, Osaka University, 
   Toyonaka, Osaka 560-0043 Japan
}

\abstract{
We report on our study of flavor-singlet mesons 
in three-flavor QCD with dynamical overlap quarks. 
Gauge ensembles are generated on a $16^3 \times 48$ lattice 
at a lattice spacing of 0.10 fm
with the strange quark masses around 
its physical value $m_{s,\rm phys}$ and up and down quark 
masses down to $m_{s, \rm phys}/5$. 
Connected and disconnected meson correlators are 
calculated using the all-to-all quark propagator.
We present our preliminary results on the spectrum 
of flavor-singlet pseudoscalar and vector mesons.
}

\FullConference{
   The XXVII International Symposium on Lattice Field Theory - LAT2009\\
   July 26-31 2009\\
   Peking University, Beijing, China
}

\begin{document}


\section{Introduction}

A quantitative understanding of interesting properties of 
the flavor-singlet mesons is an important subject in lattice QCD.
The famous $U(1)$ problem 
is a long-standing issue albeit past ceaseless efforts.
It is also well-known that 
the flavor-singlet and octet vector mesons mix with each other almost ideally,
though this has not yet been confirmed from first principles.

In this article, 
we report on our study of the flavor-singlet 
pseudoscalar (PS) and vector mesons.
There are two salient features of this work:
i) we simulate $N_f\!=\!2\!+\!1$ QCD 
including the effects of dynamical strange quarks,
which have been often ignored in previous studies
of the flavor-singlet mesons,
and ii) 
we use the all-to-all quark propagator \cite{A2A}
to calculate disconnected meson correlators,
which induce the meson mixings and 
the mass splittings from flavor-non-singlet mesons.


\section{Simulation method}


Our gauge configurations of $N_f\!=\!2\!+\!1$ QCD are 
generated on a $16^3 \times 48$ lattice 
using the Iwasaki gauge action and the overlap quark action.
We also introduce a topology fixing term \cite{exW+extmW:JLQCD} 
into our lattice action to reduce the computational cost,
and simulate only the trivial topological sector $Q\!=\!0$ 
at this stage.
The lattice spacing determined from $F_\pi$ is 0.100(5)~fm.
We take four values of the degenerate up and down quark masses
$m_l\!=\!0.015$, 0.025, 0.035 and 0.050,
which cover a range of the pion mass from 350 to 610~MeV.
Two values $m_s\!=\!0.080$ and 0.100
are chosen for the strange quark mass.
The physical quark masses fixed from $M_\pi$ and $M_K$ 
are $m_{l,\rm phys}=0.002$ and $m_{s,\rm phys}=0.065$.  
Statistics are 2,500 HMC trajectories at each combination of $m_l$ and $m_s$.
We refer readers to Ref.\cite{Nf3:Prod_Run:JLQCD} 
for further details on our gauge configurations.


We measure PS and vector meson correlators 
using the all-to-all quark propagator.
For each configuration,
we prepare 160 low-lying modes 
$(\lambda^{(k)}_m,\,u^{(k)})$ $(k\!=\!1,...,N_e(\!=\!160))$ 
of the overlap-Dirac operator $D(m)$, 
where $m$ is the valence quark mass.
Their contribution to the quark propagator is calculated exactly.
The higher modes are taken into account stochastically
by the noise method.
We prepare a single noise vector $\eta$ for each configuration, and 
dilute~\cite{A2A} it into $N_d = 3 \times 4 \times N_t/2$ vectors 
$\eta^{(d)}$ $(d=1,..,N_d)$,
which have nonzero elements 
for a single combination of color and spinor indices 
and at two consecutive time-slices.
The all-to-all propagator can be expressed as 
\bea
   D(m)^{-1}
   & = & 
   \sum_{k=1}^{N_v} v^{(k)}_m\,w^{(k)\dagger}
   \hspace{5mm}
   (N_v=N_e+N_d)
   \label{eqn:meas:a2a_prop}
\eea
with two set of vectors 
\bea
   v^{(k)}_m
   = 
   \left\{
      \frac{u^{(1)}}{\lambda^{(1)}_m}, 
      \ldots, 
      \frac{u^{(N_e)}}{\lambda^{(N_e)}_m},
      x^{(1)}_m, 
      \ldots, 
      x^{(N_d)}_m
   \right\},
   &&
   w^{(k)}
   = 
   \left\{
      u^{(1)}, \ldots, u^{(N_e)},
      \eta^{(1)}, \ldots, \eta^{(N_d)}
   \right\},
   \label{eqn:meas:vw_vectors:vw}
\eea
where $x^{(d)}_m$ is the solution of the linear equation
\bea
   D(m)\, x^{(d)}_m
   & = &
   (1-\sum_k u^{(k)}\,u^{(k)\dagger})\,\eta^{(d)}.
   \label{eqn:meas:lin_eq}
\eea

\begin{figure}[t]
\begin{center}
\includegraphics[angle=0,width=0.35\linewidth,clip]%
                {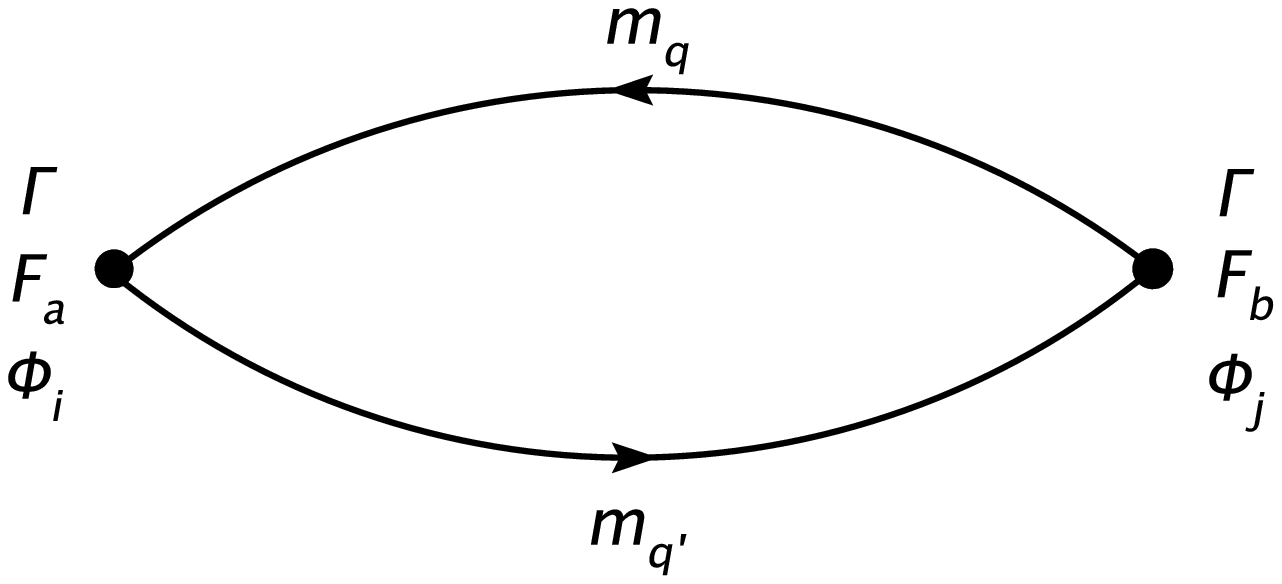}
\hspace{20mm}
\includegraphics[angle=0,width=0.35\linewidth,clip]%
                {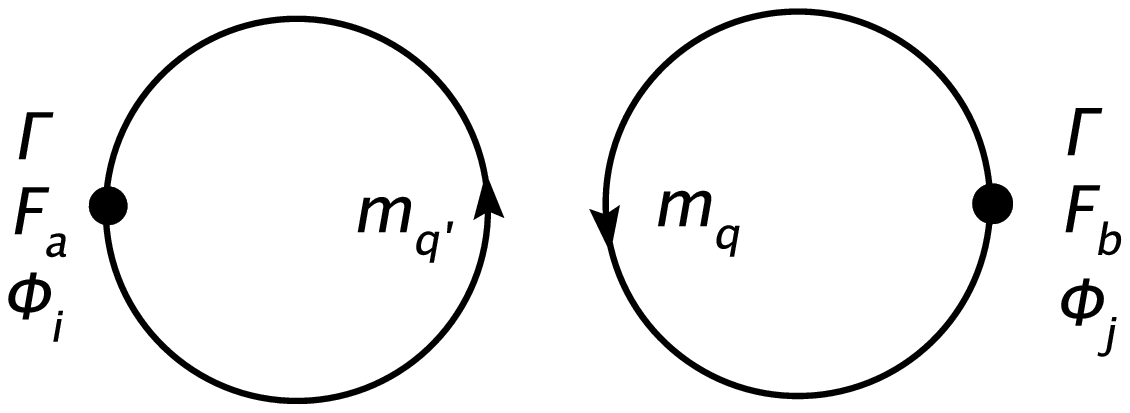}
\vspace{-2mm}
\caption{
   Connected (left diagram) and disconnected meson correlators (right diagram).
   We denote the flavor matrix and the smearing function
   for the sink meson operator by $F_a$ and $\phi_i$,
   and those for the source operator by $F_b$ and $\phi_j$.
   The Dirac matrix is $\Gamma\!=\!\gamma_5$ for PS mesons,
   whereas we average over $\Gamma\!=\!\gamma_{1,2,3}$
   for vector mesons.
   The masses of propagating quarks are denoted by $m_{q^{(\prime)}}$.
}
\label{fig:meas:corr_2pt}
\vspace{-5mm}
\end{center}
\end{figure}

We then construct the following meson field
at the temporal coordinate $t$ 
with the Dirac matrix $\Gamma$ and a smearing function $\phi(\bfr)$
\bea
   \calO^{(k,l)}_{\Gamma,\phi}(m;t)
   & = & 
   \sum_{\bfx,\bfr}
   \phi(\bfr)\, 
   w(\bfx+\bfr,t)^{(k)\dagger} \, 
   \Gamma \,
   v_m(\bfx,t)^{(l)}.
   \label{eqn:meas:meson_op}
\eea
The connected and disconnected correlators 
shown in Fig.~\ref{fig:meas:corr_2pt} 
can be calculated from these meson fields.
The PS meson correlators, for instance, are given by
\bea
    C_{P, ab, ij}(\Delta t)
    & = & 
       \frac{1}{N_t} \sum_{t}
       \sum_{q,q^\prime=u,d,s}
       \sum_{k,l=1}^{N_v}
          (F_a)_{q^\prime,q} (F_b)_{q,q^\prime} 
          \calO^{(l,k)}_{\gamma_5,\phi_i}(m_q;t+\Delta t)
          \calO^{(k,l)}_{\gamma_5,\phi_j}(m_{q^\prime};t),
    \label{eqn:meas:corr_2pt:conn}
    \\ 
    D_{P, ab, ij}(\Delta t)
    & = & 
       \frac{1}{N_t} \sum_{t}
       \sum_{q^\prime=u,d,s}
       \sum_{l=1}^{N_v}
          (F_a)_{q^\prime,q^\prime} 
          \calO^{(l,l)}_{\gamma_5,\phi_i}(m_{q^\prime};t+\Delta t)
       \sum_{q=u,d,s}
       \sum_{k=1}^{N_v}
          (F_b)_{q,q} 
          \calO^{(k,k)}_{\gamma_5,\phi_j}(m_q;t).
    \label{eqn:meas:corr_2pt:disc}
\eea
For simplicity,
we often suppress the indices of the smearing functions 
($i$ and $j$) in the following.

In this study,
we consider the PS and vector mesons in two different flavor bases:
i) 
light and strange mesons, $P_{l,s}$ and $V_{l,s}$,
with their flavor matrices
$F_l\!=\!(1/\sqrt{2}){\rm diag}[1,1,0]$ and 
$F_s\!=\!{\rm diag}[0,0,1]$,
and 
ii)
octet and singlet mesons, $P_{8,0}$ and $V_{8,0}$,
with $U(3)$ generators $T_{8,0}$ for the flavor matrices 
($F_8\!=\!T_8$ and $F_0\!=\!T_0$).
We refer to these bases as 
the light-strange and octet-singlet bases, respectively.
The full correlator of these mesons 
including the disconnected contribution is given by
\bea
   G_{\{P,V\}, ab, ij}(\Delta t)
   & = & 
   C_{\{P,V\}, ab, ij}(\Delta t)
  -D_{\{P,V\}, ab, ij}(\Delta t)
   \hspace{3mm}
   (a,b \in \{l,s\} \mbox{ or } a,b \in \{8,0\}).
\eea
We calculate all possible correlators 
with the following five different choices of the smearing function
(namely $i,j\!=\!0,...,4$)
\bea
   \begin{array}{lll}
      \phi_0(\bfr) = \delta_{\bfr,{\bf 0}},
      \hspace{20mm} 
      &
      \phi_1(\bfr) \propto \exp[-0.4|\bfr|],
      \hspace{8mm} 
      &
      \phi_2(\bfr) \propto |\bfr| \exp[-0.4|\bfr|],
      \\[1mm]
      \phi_3(\bfr) \propto \exp[-1.0|\bfr|],
      &
      \phi_4(\bfr) = \mbox{constant}
      \label{eqn:meas:smr_func}
      \\
   \end{array}
\eea
with the normalization $\sum_{\bfr} |\phi_i(\bfr)|^2 \!=\! 1$.
The calculation of all these meson correlators 
is computationally cheap,
once we prepare the $v$ and $w$ vectors of Eq.~(\ref{eqn:meas:vw_vectors:vw}).

Since the quark propagator is decomposed into low- and high-mode contributions,
the disconnected correlators can be divided into four contributions,
{\it i.e.} $D=D^{LL}+D^{LH}+D^{HL}+D^{HH}$.
We calculate these four contributions separately in our measurement.


\section{Meson correlators}

\begin{figure}[t]
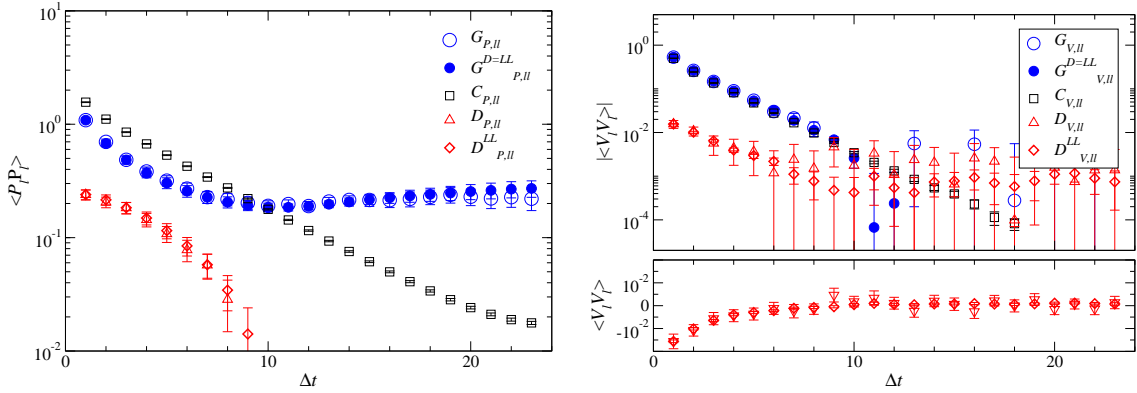

\begin{center}
\includegraphics[angle=0,width=0.48\linewidth,clip]%
                {msn_pp-sngl_mud1_ms6_mval11_smr11.eps}
\hspace{3mm}
\includegraphics[angle=0,width=0.48\linewidth,clip]%
                {msn_vv-sngl_mud1_ms6_mval11_smr22.eps}
\vspace{-2mm}
\caption{
   Light PS (left panel) and vector meson correlators (right panel)
   at $(m_l,m_s)\!=\!(0.025,0.080)$ with exponential smearing $\phi_1$ 
   for source and sink.
   Open circles, squares and triangles 
   show the full ($G_{\{P,V\},ll}$), connected ($C_{\{P,V\},ll}$) 
   and disconnected correlators ($D_{\{P,V\},ll}$), respectively.
   We also plot the low-mode contribution to the disconnected piece 
   $D_{\{P,V\},ll}^{LL}$ by diamonds, 
   and the full correlator $G_{\{P,V\},ll}^{D=LL}$ 
   defined in Eq.~(\protect\ref{eqn:corr:disc-LL}) by filled circles.  
   We note that the full PS meson correlator has a constant term
   as discussed in Section~\protect\ref{sec:ps}.
}
\label{fig:corr:ps+v}
\vspace{-5mm}
\end{center}
\end{figure}

In Fig.~\ref{fig:corr:ps+v},
we show an example of the light PS and vector meson correlators.
%
We observe that, at relatively small $\Delta t$,
the disconnected piece $D_{P,ll}$ in the full correlator $G_{P,ll}$ 
is not large
and 
it is dominated by the low-mode contribution $D_{P,ll}^{LL}$.
Therefore, 
we may safely ignore the high mode contributions to $D_{P,ll}$,
namely $D_{P,ll}^{LH}$, $D_{P,ll}^{HL}$ and $D_{P,ll}^{HH}$,
to calculate the full correlator as 
\bea
   G_{P,ab,ij}^{D=LL}(\Delta t)
   & = & 
   C_{P, ab, ij}(\Delta t)
  -D_{P, ab, ij}^{LL}(\Delta t).
   \label{eqn:corr:disc-LL}
\eea
Figure~\ref{fig:corr:ps+v} actually shows that 
$G_{P,ll}$ is well approximated by $G_{P,ll}^{D=LL}$
in the whole region of $\Delta t$.


As shown in the same figure,
the vector meson full correlator $G_{V,ll}$ turns out 
to be noisy at relatively large $\Delta t$. 
The large uncertainty mainly comes from 
those of high-mode contributions $D_{V,ll}^{\{LH,HL,HH\}}$
due to the noise method with the small number of noise samples.
We observe that 
$G_{V,ll}$ at small $\Delta t$ is well approximated by $G_{V,ll}^{D=LL}$
defined as in Eq.~(\ref{eqn:corr:disc-LL}),
and expect that the high-mode contributions $D_{V,ll}^{\{LH,HL,HH\}}$
remain to be small at larger $\Delta t$ 
since they mainly describe short distance physics.
Then $G_{V,ll}$ is expected to be well approximated by $G_{V,ll}^{D=LL}$ 
also at large $\Delta t$.

From these observations,
we use the meson correlators $G_{\{P,V\},ab}^{D=LL}$
ignoring the noisy contributions $D_{\{P,V\},ab}^{\{LH,HL,HH\}}$
to study the spectrum of the flavor-singlet mesons.
The superscript ``$D\!=\!LL$`` is suppressed in the following 
for simplicity.


\section{Vector mesons}

\begin{figure}[t]
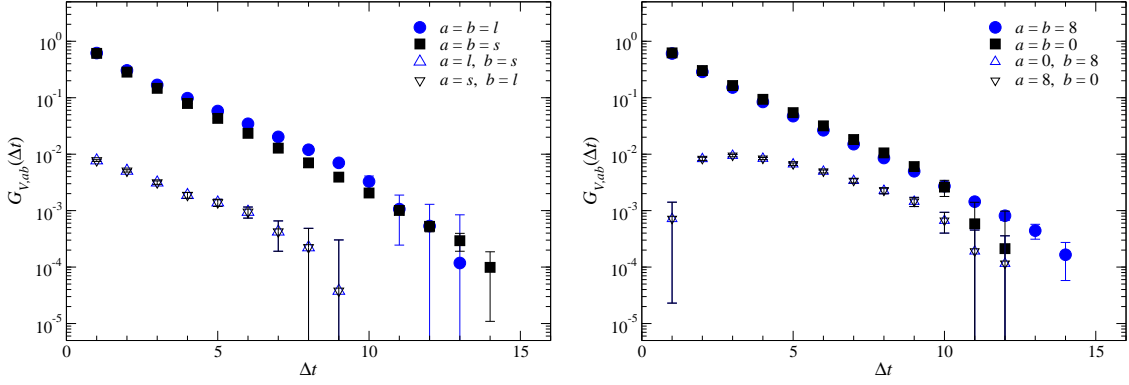

\begin{center}
\includegraphics[angle=0,width=0.48\linewidth,clip]%
                {msn_vv-sngl_mud1_ms6_smr11.eps}
\hspace{2mm}
\includegraphics[angle=0,width=0.48\linewidth,clip]%
                {msn_vv-su3_mud1_ms6_smr11.eps}
\vspace{-2mm}
\caption{
   Vector meson correlators in the light-strange (left panel)
   and octet-singlet bases (right panel).
   Both panels show 
   results with the exponential smearing function $\phi_1$
   at $(m_l,m_s)\!=\!(0.025,0.080)$.
   Filled and open symbols represent
   diagonal and off-diagonal correlators, respectively.
}
\label{fig:v:corr}
\vspace{-5mm}
\end{center}
\end{figure}

We plot the vector meson correlators in Fig.~\ref{fig:v:corr}.
In the light-strange basis, 
the off-diagonal correlators $G_{V,\{ls,sl\}}$ are 
about two orders of magnitude smaller than 
the diagonal ones $G_{V,\{ll,ss\}}$.
There is no such large hierarchy in 
$G_{V,\{88,00,80,08\}}$ in the octet-singlet basis.
Since the off-diagonal correlators induce the meson mixing,
the above observations indicate that
the mixing of the vector mesons is close to the ideal mixing:
namely, $V_8$ and $V_0$ mesons mix significantly with each other 
to form $\omega$ and $\phi$ mesons,
which  are well approximated by $V_l$ and $V_s$.

For a more quantitative examination,
we solve the generalized eigenvalue problem (GEVP)
\bea
   C(\Delta t^\prime)^{-1/2} C(\Delta t) C(\Delta t^\prime)^{-1/2} \tilde{u}_n
   & = &
   \tilde{\lambda}_n\,\tilde{u}_n
   \hspace{3mm}
   (n=0,1),
   \label{eqn:gep}
\eea
where $C(\Delta t)$ is $2 \times 2$ correlator matrix 
with specified smearing functions for source and sink
\bea
   C(\Delta t)
   & = &
   \left(
      \begin{array}{ll}
         G_{V,88}(\Delta t) & G_{V,80}(\Delta t) \\
         G_{V,08}(\Delta t) & G_{V,00}(\Delta t) \\
      \end{array}
   \right).
   \label{eqn:cm}
\eea
The creation operators of the energy eigenstates,
namely, $\phi$ and $\omega$ mesons, 
are determined from the eigenvectors $\tilde{u}_n$.
We obtain the following relation for the local operators
\bea
   \left\{
   \begin{array}{ll}
      \phi   \ = \ 0.84(5)\,V_8 - 0.55(7)\,V_0
             \ = \ 1.00(1)\,V_s - 0.04(9)\,V_l
      \\
      \omega \ = \ 0.55(7)\,V_8 + 0.84(5)\,V_0
             \ = \ 0.04(9)\,V_s + 1.00(1)\,V_l
   \end{array}
   \right.,
\eea
which implies the ideal mixing of the vector mesons.
We observe that 
this relation of vector meson operators does not change significantly
with other choices of the smearing functions.

\begin{figure}[b]
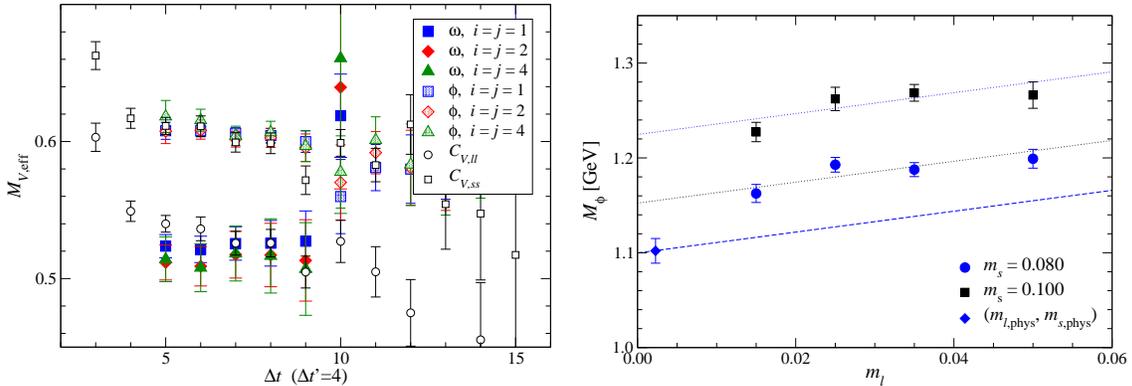

\begin{center}
\includegraphics[angle=0,width=0.48\linewidth,clip]%
                {em_msn_omega_phi_mud1_ms6.eps}
\hspace{2mm}
\includegraphics[angle=0,width=0.48\linewidth,clip]%
                {Mphi_vs_ml.eps}
\vspace{-2mm}
\caption{
   Left panel: 
   effective masses 
   of $\omega$ (solid symbols) and $\phi$ mesons (shaded symbols)
   at $(m_{ud},m_s)\!=\!(0.025,0.080)$.
   We also plot effective masses of $C_{V,ll}$ (open circles) 
   and $C_{V,ss}$ (open squares).
   Right panel: 
   chiral extrapolation of $M_\phi$.
   Dotted and dashed lines show fit lines at $m_s$ and $m_{s,\rm phys}$,
   respectively.
}
\label{fig:v:eff_mass+chiral_fit}
\vspace{-5mm}
\end{center}
\end{figure}


In Fig.~\ref{fig:v:eff_mass+chiral_fit}, 
we plot the effective masses of $\omega$ and $\phi$ mesons 
determined from the eigenvalues
$\tilde{\lambda}_n \! = \! \exp[-E_n(\Delta t - \Delta t^\prime)]$.
By taking a sufficiently large $\Delta t^\prime$,
the effective masses show 
small dependence on $\Delta t$ as well as on the smearing functions.
The same figure also shows that 
$M_{\omega}$ and $M_{\phi}$ are close to 
those from the connected correlators 
of the light and strange mesons $C_{V,\{ll,ss\}}$.
This is because 
the diagonal disconnected correlators $D_{V,\{ll,ss\}}$ 
are small as seen in Fig.~\ref{fig:corr:ps+v},
and hence they have small effects to $M_{\omega}$ and $M_{\phi}$.


In this analysis,
we extrapolate $M_{\omega(\phi)}$ to the physical point 
using a simple linear form
\bea
   M_{\omega(\phi)}
   & = &
   a_{\omega(\phi)} + b_{\omega(\phi)}m_l + c_{\omega(\phi)}m_s.
   \label{eqn:v:chiral_fit}
\eea
As shown in Fig.~\ref{fig:v:eff_mass+chiral_fit},
this fit describes our data reasonably well with 
$\chi^2/{\rm d.o.f}\!\sim\!1.3$.
We obtain 
$M_\omega = 909(25)_{\rm stat}(-98)_{\rm sys}~\mbox{MeV}$ 
and $M_\phi   = 1102(13)_{\rm stat}(-97)_{\rm sys}~\mbox{MeV}$,
where 
the systematic error is estimated by including a higher order term
$d_{\omega(\phi)} m_l^{3/2}$ \cite{ChPT:V}
and by using a different input $M_\Omega$
to fix the lattice spacing.
These results are consistent with the experimental values 
$M_\omega\!=\!783$~MeV and $M_\phi\!=\!1019$~MeV.
Note, however, that 
our data may suffer from 
significant finite volume corrections
at two smallest quark masses $m_l\!=\!0.015$ and 0.025,
where $2.8 \! \leq \! M_\pi\,L \! \leq \! 3.2$.
We are planning to extend this work to a larger volume $24^3 \times 48$
for a more precise comparison with the experiment.


\section{PS mesons}
\label{sec:ps}


Figure~\ref{fig:ps:corr} shows 
correlators of the light and strange PS mesons.
In contrast to the vector mesons,
the off-diagonal correlators $G_{P,\{ls,sl\}}$
are not so small compared to the diagonal ones $G_{P,\{ll,ss\}}$
in the light-strange basis.
This leads to a significant strange (light) quark component
in $\eta$ ($\eta^\prime$).
For the local operators, we obtain 
\bea
   \eta        = 0.96(1)\,P_l - 0.28(3)\,P_s,
   \hspace{3mm}
   \eta^\prime = 0.28(3)\,P_l - 0.96(1)\,P_s.
   \label{eqn:ps:mix}
\eea
As in phenomenological analyses \cite{PS_mixing},
more unambiguous determination of the mixing matrix could be provided 
by constructing the local $\eta$ and $\eta^\prime$ operators 
so that their decay constants reproduce the experimental values.
We leave this for a future study.

\FIGURE{
   \centering 
   \includegraphics[angle=0,width=0.48\linewidth,clip]%
                   {msn_pp-sngl_mud1_ms6_smr11.eps}
   \vspace{-4mm}
   \caption{ 
      Light and strange PS meson correlators.
   }
   \label{fig:ps:corr}
   \vspace{0mm}
}


As predicted analytically \cite{fixedQ}
and as seen in Figs.~\ref{fig:corr:ps+v} and \ref{fig:ps:corr},
disconnected contribution $D_{P,ab}$ induces 
a constant term in the full correlator $G_{P,ab}$ at fixed topology, 
{\it e.g.}
\bea
   m_l G_{P,ll}
   \xrightarrow[\Delta t \to \infty]{} 
   \frac{\chi_t}{V}\left( 
                      1 - \frac{Q^2}{\chi_t V} + \frac{c_4}{2 \chi_t^2 V}
                   \right),
   &&
   \hspace{5mm}
   \label{eqn:ps:fixedQ}
\eea
which is suppressed by $1/V$. 
While this term is useful 
to determine the topological susceptibility $\chi_t$ \cite{Nf2:chi_t:JLQCD},
this forces us to use $G_{P,ab}$ at small $\Delta t$
to extract the PS meson masses $M_\eta$ and $M_\eta^\prime$.

To eliminate excited state contamination at such small $\Delta t$,
we solve the GEVP with the $10 \times 10$ correlator matrix 
including the smearing degrees of freedom
\bea
   C(\Delta t) 
   & = &
   \left(
      \begin{array}{lll}
         G_{P,88,00}(\Delta t) & \cdots &
         G_{P,80,04}(\Delta t)             \\
         \cdots & \cdots & \cdots                     \\
         G_{P,08,40}(\Delta t) & \cdots &
         G_{P,00,44}(\Delta t)             \\
      \end{array}
   \right).
\eea
Effective masses of $\eta$ and $\eta^\prime$ mesons
are plotted in Fig.~\ref{fig:ps:eff_mass}.
Although $\eta$ seems to be lighter than $\eta^\prime$,
the existence of the constant term in Eq.~(\ref{eqn:ps:fixedQ})
leads to a large uncertainty of $M_{\eta^\prime}$: 
10\,--\,15\% already at simulated quark masses.


From a linear chiral extrapolation in terms of $m_l$ and $m_s$, 
we obtain 
$M_\eta \! = \! 639(50)_{\rm stat}~\mbox{MeV}$ 
and $M_{\eta^\prime} = 840(136)_{\rm stat}~\mbox{MeV}$ 
at the physical point.
These are consistent with the experimental values
$M_\eta \! = \! 548$~MeV and $M_{\eta^\prime} \! = \! 958$~MeV,
though 
the statistical significance of the $\eta^\prime$\,--\,$\eta$ 
mass splitting (1.4\,$\sigma$) is not sufficient.


\section{Conclusion}

\FIGURE{
   \centering 

   \vspace{0mm}
   \includegraphics[angle=0,width=0.48\linewidth,clip]%
                   {em_msn_eta_etap_mud1_ms6.eps}
   \vspace{-2mm}
   \caption{
      Effective masses of $\eta^\prime$ (top panel) 
      and $\eta$ (bottom panel)
      at $(m_{ud},m_s)\!=\!(0.025,0.080)$
      with different choices of $\Delta t^\prime$.
      For a comparison, 
      we also plot 
      effective mass from the connected strange correlator $C_{P,ss}$.
   }
   \label{fig:ps:eff_mass}
}

In this article,
we report on our study of the flavor-singlet mesons
in $N_f\!=\!2+1$ lattice QCD
using the all-to-all quark propagator 
to calculate the disconnected correlators.
For the vector meson,
we observe that the small disconnected contributions 
in the light-strange basis
lead to the almost ideal mixing and the small mass shift.
This is consistent with the experimental fact
$M_\omega \! \sim \! M_\rho$,
and explains why the previous calculations of $M_\phi$ 
ignoring the disconnected contributions 
show reasonable agreement with experiment 
\cite{Review:Spectrum:Lat09}.

We need to improve the accuracy of $M_{\eta^\prime}$
to establish the $\eta^\prime$\,--\,$\eta$ mass splitting.
This could be done by simulating non-trivial topological sectors
as well as 
by suppressing the fixed topology effects in Eq.~(\ref{eqn:ps:fixedQ})
on a larger lattice.
The latter is also important 
to suppress finite volume corrections to the PS and vector meson masses 
for a more detailed comparison with the experiment.
Such simulations on a $24^3 \times 48$ lattice are in progress.

\vspace{5mm} 

Numerical simulations are performed 
on Hitachi SR11000 and IBM System Blue Gene Solution 
at High Energy Accelerator Research Organization (KEK)
under a support of its Large Scale Simulation Program (No.~09-05).
This work is supported in part by the Grant-in-Aid of the
Ministry of Education 
(No.~19740121, 20105001, 20105002, 20105003, 20105005, 20340047, 21105508, 
21674002 and 21684013),
the National Science Council of Taiwan
(No.~NSC96-2112-M-002-020-MY3, NSC96-2112-M-001-017-MY3, NSC98-2119-M-002-001),
and NTU-CQSE (No.~97R0066-65 and 97R0066-69).
The work of HF was supported by the Global COE program of Nagoya
University "QFPU" from JSPS and MEXT of Japan.


\end{document}